\documentclass[%
 reprint,
superscriptaddress,
 amsmath,amssymb,
 aps,
]{revtex4-2}

\usepackage{graphicx}% Include figure files
\usepackage{dcolumn}% Align table columns on decimal point
\usepackage{bm}% bold math
\usepackage{color}
\usepackage{comment}

\def\red#1{\textcolor{red}{#1}}
\def\e{\begin{equation}}
\def\f{\end{equation}}
\def\=#1{\overline{\overline #1}}

\def\E{\varepsilon}

\def\Re{{\rm Re\mit}}

\def\l#1{\label{eq:#1}}
\def\rr#1{(\ref{eq:#1})}
\def\_#1{{\bf #1\mit}}

\begin{document}

\preprint{APS/123-QED}

\title{Sub-Wavelength Passive Optical Isolators Using Photonic Structures Based on Weyl Semimetals} 

% \title{Wavelength-Scale Passive Optical Isolators Using Weyl Semimetals} 

% \title{Magnetic Weyl Semimetal Optical Isolator}% Force line breaks with \\

\author{Viktar Asadchy}\thanks{These authors contributed equally.}
\affiliation{%
 Ginzton Laboratory and Department of Electrical Engineering, Stanford University, Stanford, California 94305, USA
}

\author{Cheng Guo} \thanks{These authors contributed equally.}
 \affiliation{Department of Applied Physics, Stanford University, Stanford, California 94305, USA}
 
\author{Bo Zhao}
\affiliation{%
 Ginzton Laboratory and Department of Electrical Engineering, Stanford University, Stanford, California 94305, USA
} 

\author{Shanhui Fan}
\email{shanhui@stanford.edu}
\affiliation{%
 Ginzton Laboratory and Department of Electrical Engineering, Stanford University, Stanford, California 94305, USA
}

%              %  but any date may be explicitly specified

\begin{abstract}
We design sub-wavelength high-performing non-reciprocal optical devices using recently discovered magnetic Weyl semimetals. These passive bulk topological materials exhibit anomalous Hall effect which results in magnetooptical effects that are orders of magnitude higher than those in conventional materials, without the need of any  external  magnetic  bias.  We design two optical isolators of both Faraday and Voigt geometries. These isolators have dimensions that are reduced by three orders of  magnitude compared to conventional magneto-optical configurations.
Our results indicate that the magnetic Weyl semimetals may open up new avenues in photonics for the design of various nonreciprocal components.
\end{abstract}

\maketitle
Isolators are nonreciprocal devices that allow light propagation in one direction but block it in the opposite direction. They play a fundamental role in modern communication systems, being used, for example, for suppressing back reflection in lasers and reducing multipath interference in communication channels. The ultimate requirements for an ideal isolator device are high isolation, compact size, low insertion loss, large bandwidth, and the absence of active components and external magnetic field.  

Isolation inherently requires breaking Lorentz reciprocity, which can be achieved by breaking  time-reversal symmetry or exploiting nonlinear response~\cite[Sect.~XII]{landau_course_1980},\cite{jalas_what_2013,caloz_electromagnetic_2018}. 
%\cite{tutorial}
Traditionally,  isolators are designed based on magneto-optical materials such as yttrium iron  or terbium gallium garnets, biased by an external magnetic field which breaks time reversal symmetry~\cite{zvezdin_modern_1997}. Ferrite-based isolators at microwave frequencies typically have  dimensions of the operational wavelength or smaller due to strong gyromagnetic
resonances~\cite[\textsection~9.4]{pozar_microwave_2012}. On the contrary, at optical frequencies nonreciprocity in ferrite results from the  nonresonant (weak) electron cyclotron orbiting, yielding optical isolators designs with dimensions of thousands of the operational wavelength or larger~\cite[\textsection~13.3]{zvezdin_modern_1997}. 
The size can be partially decreased using geometries based on ring resonators or Mach-Zehnder interferometers, however, it is still of the order of millimeters~\cite{du_monolithic_2018,zhang_monolithic_2019}. Another alternative for the size reduction  of magneto-optical isolators is the use of magnetic photonic crystals~\cite{figotin_nonreciprocal_2001,lyubchanskii_magnetic_2003,yu2007,khanikaev_low-symmetry_2009,fang_ultracompact_2011}. Nevertheless, they require either strong and nonuniform magnetization that is difficult to realize, or narrow-band resonances.  

As modern photonics requires ultimate miniaturization of optical components, in the last decades, significant efforts have been devoted to the design of compact optical isolators based on time-modulated~\cite{cullen_travelling-wave_1958,yu_complete_2009,lira_electrically_2012,fang_photonic_2012,Sounas_time_modulation_2017} and nonlinear~\cite{tocci_thinfilm_1995,peng_parity_2014,fan_all-silicon_2012,shi_limitations_2015} systems. Additional advantage of these systems over conventional magneto-optical materials is that they do not require an external magnetic field, or generate a static magnetic field through their intrinsic magnetic moment. These magnetic fields could negatively influence the operation of other optical components located in the closed proximity to the isolator. Nevertheless, practical realization of time-modulated isolators remain very challenging since it requires active sources, while nonlinear isolators have intrinsic limitations on their operations due to dynamic reciprocity~\cite{shi_limitations_2015}. 

In this Letter, we demonstrate that gigantic optical isolation can be achieved by designing photonic structures based on  recently discovered Weyl semimetals~\cite{Hosur2013,yan2017,armitage2018a,belopolski2019,morali2019,liu2019}. It is known that these bulk topological materials can  exhibit simultaneously broadband and giant magnetooptical effect without an external magnetic bias. Moreover,     Weyl semimetals can be antiferromagnetic, exhibiting zero magnetic moment~\cite{yang_topological_2017,zhang_strong_2017}, and hence does not generate an external magnetic field. Based on the giant optical nonreciprocity in Weyl semimetals, we propose two designs of   isolators at  mid-infrared wavelengths with Faraday and Voigt geometries, respectively. The Faraday isolator provides more than 40~dB isolation and 0.33~dB insertion loss with unprecedented sub-wavelength thickness of a quarter wavelength (excluding polarizers and antireflective coatings). The Voigt isolator provides around  31.5~dB isolation and 1.2~dB insertion loss with a thickness of 1.38 wavelength, without the need of additional polarizers and antireflective coatings. These results enable creation of a new generation of passive optical isolators with dimensions reduced by three orders of magnitude compared to the previous solutions.

%In this Letter, we demonstrate that gigantic optical isolation can be achieved using recently discovered Weyl semimetals~\cite{Hosur2013,yan2017,armitage2018a,belopolski2019,morali2019,liu2019}. We show that these bulk topological materials exhibit the record-breaking Verdet  coefficient characterizing the magnetic circular birefringence  without any external magnetic bias (only due to the internal magnetic ordering of the crystal structure). More importantly, theoretically Weyl semimetals do not necessary generate a net magnetic field outside (avoiding interference on possible optical components near the isolator), as suggested by the  discovered Weyl semimetal phase in  chiral antiferromagnetic
%materials such as  ${\rm Mn}_3{\rm Sn}$~\cite{yang_topological_2017,zhang_strong_2017}. 
%We  design an optical isolator based on bulk Weyl semimetal where separation of the Weyl nodes in the momentum space is parallel to the direction of incident waves (equivalent to the Faraday configuration for magneto-optical materials). Designed for mid-infrared, it provides more than 40~dB isolation and 0.33~dB insertion loss with unprecedented sub-wavelength thickness of a quarter wavelength.  Furthermore, we report giant optical isolation without polarizers in a one-dimensional photonic crystal based on Weyl semimetals with the momentum separation perpendicular to the incidence direction. These results enable creation of a new generation of magnetless passive optical isolators with dimensions reduced by three orders of magnitude compared to the previous solutions.

Weyl semimetals are a novel class of three-dimensional gapless topological matter~\cite{Xu2015,Lv2015,belopolski2019,morali2019,liu2019,kuroda2017,hirschberger2016}. They feature accidental degenerate Weyl nodes in their band structure that host chiral fermions and appear in pairs of opposite chirality~\cite{armitage2018a}. Each Weyl node acts as a source/drain of Berry curvature in   momentum space. The nontrivial topology of the Weyl nodes leads to unique electronic and optical properties that have generated significant interest in  both  fundamental science and technology. In the context of photonic applications, Weyl semimetals have been proposed recently for generation of nonreciprocal surface plasmons and design of nonreciprocal thermal emitters~\cite{kotov2018,zhao2019,tsurimaki2019a}.  %Realization of a Weyl semimetal requires breaking of either inversion ($\mathcal{P}$) or time-reversal ($\mathcal{T}$) symmetry. Noncentrosymmetric Weyl semimetals that break $\mathcal{P}$ but preserve $\mathcal{T}$ symmetry have been discovered since 2015  , but only until more recently have magnetic Weyl semimetals that break $\mathcal{T}$ been experimentally discovered   \cite{belopolski2019,morali2019,liu2019,kuroda2017,hirschberger2016}. 

\begin{comment}
The nontrivial topology of the Weyl nodes gives rise to novel electromagnetic responses that are drastically different from conventional materials. These include the chiral magnetic effect and the anomalous Hall effect~\cite{armitage2018a}. Both effects can be represented compactly by the formalism of axion electrodynamics~\cite{wilczek1987}, which adds the axion term to the electromagnetic Lagrangian density~\cite{zyuzin2012}:
\red{\begin{align}
\label{eq:lagrangian}
    \mathcal{L}_\theta &= 2\alpha \sqrt{\frac{\epsilon_0}{\mu_0}}\frac{\theta(\mathbf{r},t)}{2\pi}\mathbf{E}\cdot \mathbf{B},
\end{align}}
where  %$\mathcal{L}_0$ and $\mathcal{L}_\theta$ are the usual and axion parts of the Lagrangian density, respectively.
$\alpha = \tfrac{e^2}{4\pi\epsilon_0\hbar c}$ is the fine structure constant, $e$ is the elementary charge, $\hbar$ is the reduced Planck constant, $\epsilon_0$ is the permittivity of vacuum, $\mu_0$ is the permeability of vacuum, $\mathbf{E}$ is the electric field, $\mathbf{B}$ is the magnetic flux density, and $\theta(\mathbf{r},t)$ is the axion angle that has space and time dependence. %We use SI unit throughout the paper.}
\end{comment}

\begin{figure}[tb]
	\centering
	\includegraphics[width=0.7\linewidth]{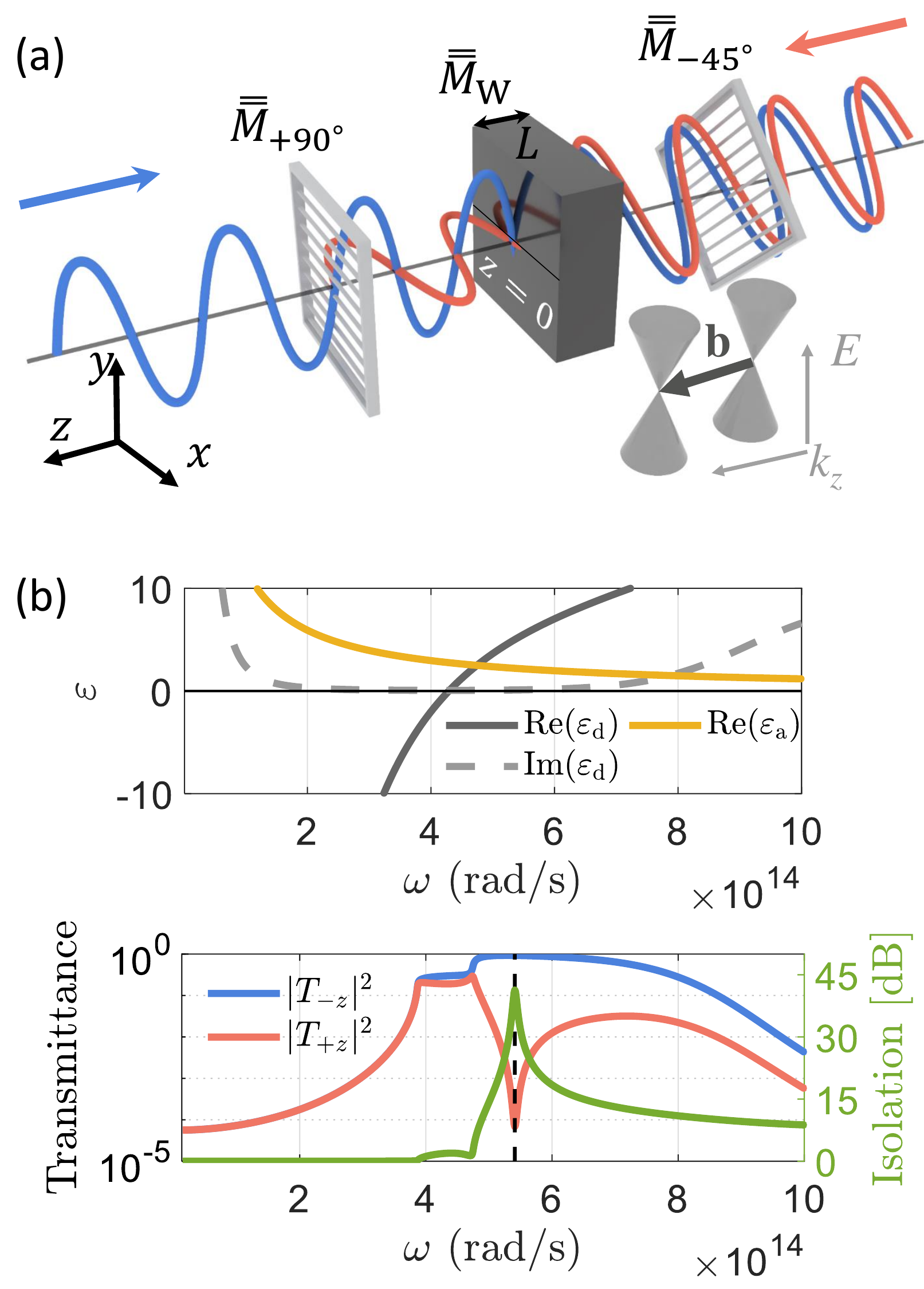} 
	\caption{	(a) Geometry of the Faraday isolator based on Weyl semimetal slab of thickness~$L$ and Weyl nodes separation in the  momentum space~$ \_b$. The inset shows  the bulk energy $E$ dispersion of the semimetal. The silver grids depict linear polarizers. The red and blue traces denote the electric fields of linearly polarized waves propagating, respectively, along the  $-z$- and $+z$-directions. Matrices $\overline{\overline{M}}_{+90^\circ}$, $\overline{\overline{M}}_{-45^\circ}$, and $\overline{\overline{M}}_{\rm W}$ stand for the Jones matrices of the two polarizers and the Weyl semimetal slab. 
	(b) Up: Angular frequency dispersion of the permittivity components of  bulk Weyl semimetal with the   parameters defined in the main text.
% 	: $\tau=1000$~fs, $b=0.085$~${\rm \AA}^{-1}$, $E_{\rm F}=0.3$~eV, $T=300$~K, $v_{\rm F} = 0.83 \times 10^5$~m/s and $g=2$.
	Bottom:  Transmittances for waves travelling through the Weyl isolator along $\pm z$-directions with polarization states shown in (a). The blue and red colors in (a) and (b) correspond to the same waves. The green curve shows the isolation ratio. The dashed line depicts the frequency of the highest isolation.    }
	\label{fig1}
\end{figure}
For concreteness, we consider the simplest case of a magnetic Weyl semimetal that hosts two Weyl nodes of opposite chirality at the same energy with a momentum separation $2\hbar\mathbf{b}$. Such an ideal Weyl semimetal phase has been experimentally realized very recently in ${\rm Eu Cd_2 As_2}$~\cite{soh2019}. For this semimetal, the electric permittivity assumes the form of  the so-called axion electrodynamics~\cite{Hofmann2016}:
\begin{equation} 
\mathbf{D} = \E_d \mathbf{E} + \frac { i e ^ { 2 } } {  4 \pi^2  \hbar \omega } 2\mathbf{b} \times \mathbf{E}, 
\l{constitutive}\end{equation}
while $\_B=\mu_0 \_H$ is not modified. Here, $e$ is the electron charge, $\hbar$ is the reduced Planck constant, $\omega$ is the angular frequency, and $\varepsilon_d$ denotes the    permittivity of the corresponding Dirac semimetal with doubly degenerate bands, which is assumed isotropic. 
The second term in Eq.~\rr{constitutive} describes the anomalous Hall effect in the dynamical regime, manifesting itself as the   magneto-optical effect~\cite{nagaosa2010}. The axial vector  $2\mathbf{b}$ acts as an effective magnetic field. Its effect on the dielectric permittivity is similar to that of an external magnetic field on a plasmonic metal. In this case, however, $\mathbf{b}$ is entirely due to the electronic structure of the Weyl semimetal, is entirely internally generated, and does not correspond to an external magnetic field. If we choose the crystal orientation of the Weyl semimetal   such that $\mathbf{b}$ is along the $\_{\hat{z}}$ direction: $\mathbf{b} = b \mathbf{\hat{z}}$, the effective permittivity tensor in $\_D = \overline{\overline{\varepsilon}} \cdot \_E $ becomes:
\begin{equation}\label{eq:epsilon_tensor}
\overline{\overline{\varepsilon}} =\begin{pmatrix}
\varepsilon_{\rm d}&i\varepsilon_{\rm a}&0\\
-i\varepsilon_{\rm a}&\varepsilon_{\rm d}&0\\
0&0&\varepsilon_{\rm d}
\end{pmatrix},
\end{equation}
where $\varepsilon _ { \rm a } = b e ^ { 2 }  /(2 \pi^2 \hbar \omega) $. 
Thus, $\overline{\overline{\varepsilon}}$ is asymmetric and breaks Lorentz reciprocity~\cite[\textsection~5.5c]{kong_electromagnetic_1986}.  
The diagonal term $\varepsilon_{ \rm  d}$ is calculated by using the Kubo-Greenwood formalism within the random phase approximation to a two-band model with spin degeneracy~\cite{Hofmann2016,kotov2016,kotov2018}:
\begin{align}
\label{eq:ed}
    \varepsilon_{\rm d} = \varepsilon _ { \rm b } &+ \frac {i r _ {\rm s } g } { 6 \omega} \Omega \, G ( \Omega / 2 ) - \frac { r _ { \rm s } g } { 6 \pi  \omega} \left\{ \frac { 4 } { \Omega } \left[ 1 + \frac { \pi ^ { 2 } } { 3 } \left( \frac { k _ {\rm B } T } { E _ { \rm F } (T)} \right) ^ { 2 } \right] \right.\notag\\
    &+ \left. 8 \Omega \int _ { 0 } ^ { \xi _ {\rm c } } \frac { G ( \xi ) - G ( \Omega / 2 ) } { \Omega ^ { 2 } - 4 \xi ^ { 2 } } \xi d \xi \right\}\, ,
\end{align}
here $\varepsilon_{\rm b}$ is the background permittivity, $E_{\rm F}$ is the chemical potential, $\Omega = \hbar(\omega+i\tau^{-1})/E_{ \rm F}$  is the normalized complex frequency, $\tau^{-1}$ is the Drude damping rate, $G(E) = n(-E)-n(E)$,  where $n(E)$ is the Fermi distribution function,   $r_{ \rm s} = e^2/4\pi\epsilon_0\hbar v_{ \rm F}$ is the effective fine-structure constant, $v_{ \rm F}$ is the Fermi velocity, $g$ is the number of Weyl points, and $\xi_{ \rm c} =E_{ \rm c}/E_{ \rm F}$, where $E_{ \rm c}$ is the cutoff energy beyond which the band dispersion is no longer linear~\cite{kotov2016}. Following Ref.~[\citenum{kotov2018}], in this work we use the parameters $\varepsilon_{ \rm b} = 6.2$, $\xi_{ \rm c} = 3$, $\tau= 1000$~fs, $g=2$, $b=8.5\times 10^8 \, { \rm m}^{-1}$, $v_{ \rm F} = 0.83 \times 10^5$~m/s, and  $E_{ \rm F} = 0.30$~eV at $T=300$~K. 

Figure~\ref{fig1}(b) (upper plot) depicts frequency dispersion of permittivity components~$\varepsilon_{ \rm d}$ and $\varepsilon_{ \rm a}$. It is seen that  $\varepsilon_{ \rm a}$ is comparable to $\varepsilon_{ \rm d}$ over a broad wavelength range. This dispersion
indicates   strong nonreciprocity in Weyl semimetals characterized with the magneto-optical parameter~\cite{zvezdin_modern_1997} $Q \equiv \varepsilon_{ \rm a}/\varepsilon_{ \rm d} \sim 1$, which is around three orders of magnitude larger than that of conventional  magneto-optical materials in the optical frequencies with $Q\sim 10^{-3}$\cite{zvezdin_modern_1997}. Such a strong nonreciprocity is induced by the intrinsic Berry curvature in Weyl semimetals. It is naturally broadband, in sharp contrast to the magnetooptical effects induced by Zeeman splitting in two level systems~\cite{ying2018}. It can appear in the absence of an external magnetic field. Moreover, the magnetic Weyl semimetal can be antiferromagnetic~\cite{yang_topological_2017,zhang_strong_2017}, i.e. exhibit zero net magnetic moment. Antiferromagnetic nonreciprocal optical devices will benefit integrated optoelectronics as they do not generate net magnetic field outside, which would otherwise interfere with other adjacent  components.

The giant nonreciprocity in magnetic Weyl semimetals results in significant magnetooptical effects, in both the Faraday geometry  where the light travels along the Weyl node separation ($\mathbf{k}\parallel \mathbf{b}$), and the Voigt geometry where the light travels perpendicularly to the Weyl node separation ($\mathbf{k}\bot \mathbf{b}$)~\cite{zvezdin_modern_1997}. Here $\mathbf{k}$ denotes the wavevector of light. 

First, we consider the Faraday geometry: For linearly polarized light propagating in the Weyl semimetal along  $\mathbf{b}$ direction, the plane of polarization is rotated by the angle~\cite[\textsection~3.1]{zvezdin_modern_1997}
\begin{equation} \theta_{\rm F} =V_{\rm W}(\omega) L= \Re (n_-- n_+)\frac{\omega }{2c}  L,
\l{faraday} \end{equation}
where $V_{\rm W}(\omega)$ is the modified Verdet coefficient defined as rotation per unit path, $L$ is the light propagation distance inside Weyl semimetal, $c$ is speed of light in vacuum, $n_+= \sqrt{\varepsilon_d-\varepsilon_a}$ and $n_- = \sqrt{\varepsilon_d+\varepsilon_a}$ are the refractive indices for right and left circularly polarized light, respectively. The angle $\theta_{\rm F}$ is counted from the $x$ axis towards the $y$ axis. 
%the magnetic circular birefringence without external magnetic field~\cite[\textsection~3.1]{zvezdin_modern_1997}. The latter effect is expressed in the fact that the refractive indices for right- and left-handed
%circularly polarized light propagating along the direction of the momentum separation~$2\_b$ (the Faraday configuration)
%in magnetic Weyl semimetal are different: $n_+^2=\E_{\rm d} -\E_{\rm a}$ and $n_-^2=\E_{\rm d} +\E_{\rm a}$, respectively. 
%This effect manifests itself in the rotation of the plane of polarization of linearly polarized light incident on Weyl semimetal  by the angle
%\e \theta_{\rm F} = \frac{\omega L}{2c} \Re (n_- - n_+)=V_{\rm W}(\omega) L,
%\l{faraday} \f
%where $c$ is speed of light, $L$ is the light propagation distance inside Weyl semimetal, and $V_{\rm W}(\omega)$ is the modified Verdet  coefficient defined as the rotation per unit path.  The angle $\theta_{\rm F}$ in our geometry is counted from the $x$ axis towards the $y$ axis.
%The modified Verdet  coefficient can be compared to the product $V(\omega) B_0$ traditionally calculated for magneto-optical materials biased by an external magnetic field $B_0$~\cite[\textsection~1.2]{zvezdin_modern_1997}. 
We have calculated the modified Verdet  coefficient for the Weyl semimetal (see Supplemental Materials~\cite[\textsection~1]{suppl}). The maximum $V_{\rm W}$ reaches $1.67~{\rm rad/\mu m}$ at $4~\mu$m. For comparison, to reach such an value at this wavelength using conventional magneto-optical material  such as  \mbox{YbBi:YIG}, %which has one of the highest reported Verdet coefficient among the conventional magneto-optical materials~\cite{zhao_magneto-optic_2001}, 
one needs to apply an  impractical external magnetic field that exceeds $300$~T. Moreover, compared to the giant Faraday rotation, the attenuation  can be moderate in Weyl semimetals. As a figure of merit, the dimensionless ratio of the Verdet coefficient and the attenuation coefficient for linearly polarized light $V_{\rm W}/\alpha\approx 13$ near wavelength of $4\,\mu$m (see Supplemental Materials~\cite[\textsection~1]{suppl}). Therefore, it is possible to design an isolator with low insertion loss based on Weyl semimetals. 

The geometry of the designed Faraday isolator based on Weyl semimetal is shown in Fig.~\ref{fig1}a. The semimetal slab of thickness~$L$ is encompassed by two linear polarizers twisted at $45^\circ$. The left interface of the slab is located at $z=0$, while the right one at $z=-L$.  
For this example, we assume that  reflections at the interface between Weyl semimetal and free space can be suppressed by adding antireflective coatings at the two sides of the slab.  
\begin{figure*}[tb]
	\centering
	\includegraphics[width=0.98\linewidth]{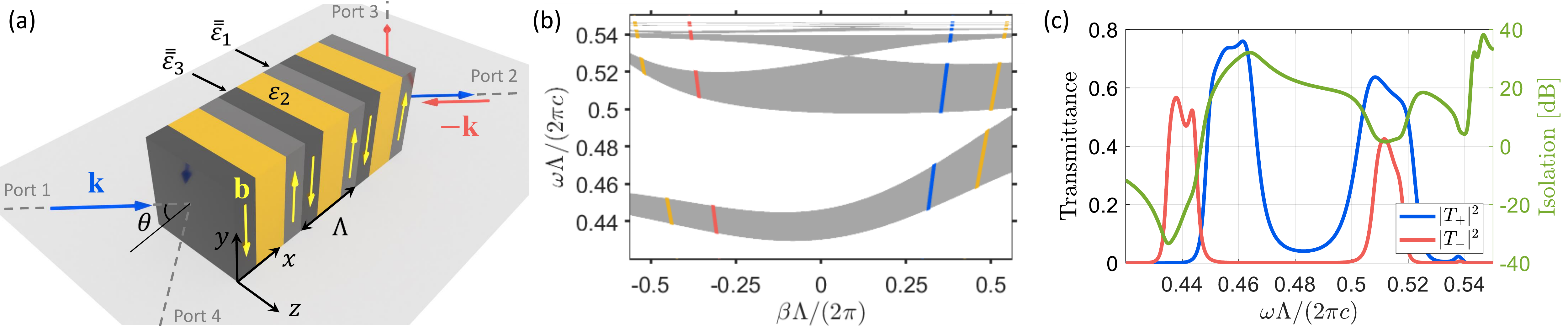} 
	\caption{(a) Geometry of the Weyl photonic crystal with the unit cell comprising three layers. The materials of the dark and bright grey layers are  Weyl  semimetals with opposite node separation shown by the yellow arrows. The dielectric films are shown in yellow. The blue and red arrows depict forward and backward illuminations, respectively. (b) Projected band  structure of an infinite Weyl photonic crystal shown in~(a). Here, $\beta$ denotes the wavevector components parallel to the layers.  The gray regions indicate the allowed bands. The yellow color denotes the light lines, while the blue and red vertical lines indicate dispersion relations for forward and backward incidence at $\theta=45.6^\circ$, respectively.  (c) Transmittances and isolation ratio for the forward and backward illuminations of the photonic crystal.  }
	\label{fig2}
\end{figure*}
Figure~\ref{fig1}b (bottom plot) depicts the calculated transmittance for both illuminations of the Weyl isolator with a thickness $L=0.886$~$\mu$m using Jones matrix analysis (See Supplemental Material ~\cite[\textsection~2]{suppl}). At an angular  frequency $\omega=5.41 \times 10^{14}$~rad/s ($\lambda=3.5~\mu$m), the isolation reaches the maximum value of 41.3~dB supported by very low insertion loss of $0.33$~dB. Importantly, such great isolation property occurs for the Weyl slab of extraordinary small thickness of $L=0.25\lambda$. For comparison, an optical isolator based on garnet \mbox{YbBi:YIG} at the same wavelength must have thickness of at least $L=150$~$\mu$m when biased by an external magnetic field $B_0=1$~T. %Another indicative comparison could be made with respect to   semiconductor indium antimonide~InSb which found to possess very strong magneto-optical effects~\cite{maack_size-dependent_2017,buddhiraju_absence_2018}. Although isolator based on InSb can be as thin as $L=0.15\lambda_{\rm THz}$ (see Ref.~\cite[\textsection~3]{suppl}), it can be designed only for terahertz frequency range ($\sim 1$~THz) and requires an external magnetic field. 
The sub-wavelength thickness of the Weyl isolator and the absence of the external bias make it very appealing candidate for future photonics applications.  

Next, we consider the Voigt geometry, which is of interest for several reasons. Firstly, the relative orientation of the Weyl node separation with respect to the crystal surface depends on the specific material and the growth condition. Thus, in some cases   vector  $\mathbf{b}$ is restricted to be parallel to the film surface. Secondly, Voigt isolators do not require additional polarizers or antireflective coating used in Faraday geometry, and hence may be simpler in implementation. Voigt isolators based on conventional magnetooptical materials have been proposed using one-dimensional photonic crystal structures~\cite{lyubchanskii_magnetic_2003,yu2007,khanikaev_low-symmetry_2009}. Due to the weak nonreciprocity in conventional magnetooptical materials, as many as 50 unit cells of the photonic crystal are needed to realize optical isolation. %We anticipate Weyl semimetals to open novel possiblities for Voigt isolator design. Due to their strong nonreciprocity, one can construct efficient and compact Voigt isolators consisting of just a few layers with a minimal total thickness.

Here we show that the use of Weyl semimetal can lead to far more compact Voigt isolators. For this purpose, we consider a finite  one-dimensional photonic crystal structure incorporating alternating layers of Weyl semimetals with $\mathbf{b}$ vector parallel to the layers and isotropic dielectric films. %In our design, we do not use birefringent materials since natural birefringence is typically weak with $|n_{\rm e} - n_{\rm o}|\sim 0.01\,$-$\,0.1$~\cite[p.~176]{fowles_introduction_1989}, which will hamper the miniaturization. 
We determine the simplest configuration of the primitive cell of the photonic crystal by symmetry considerations. 
The spectral nonreciprocity, i.e. $\omega (\_k) \neq \omega (-\_k)$, requires that the magnetic symmetry group $G$ of the structure must not include any element $g \in G$ that
flips the wavevector, i.e. $\forall g\in G, g \_k \neq -\_k$~\cite{vitebsky_electronic_1997}. It follows that for the materials we consider (Weyl semimetal and isotropic dielectric), the simplest primitive cell must consist of three layers, since with the materials that we assume here, any infinite 1D photonic crystal with a unit cell containing two layers  always maintains  spatial inversion symmetry $\mathcal{I}\_k = -\_k$. We note that a finite 1D crystal with two layers per unit cell may break inversion symmetry, and thus exhibit non-reciprocal behavior due to time-reversal symmetry breaking in the Weyl semimetal, however, the nonreciprocal effect is usually weak (see Supplemental  Materials~\cite[\textsection~3]{suppl}).
As for photonic crystals with three layers per unit cell, there are two possible configurations. The first configuration consists of two different dielectric layers and single Weyl semimetal layer in the primitive cell, which is examined in Supplemental Materials~\cite[\textsection~6]{suppl}. The second configuration consists of two Weyl layers and one dielectric layer, which exhibits the best characteristics in terms of the isolation and compactness and is considered below.

%Due to this condition, it  immediately follows that a uniform material slab possesses space inversion symmetry $\mathcal{I}$ and  is always spectrally reciprocal since $\mathcal{I} \_k =-\_k$. 
%The only exception is a slab which breaks both time- and space-reversal symmetries (see  the metasurfaces emulating such a bianisotropic uniform slab in~\cite{degiron_one-way_2014,mousavi_gyromagnetically_2014}).  
%An infinite periodic photonic crystal whose unit cell consists of only two different layers, likewise, has space inversion symmetry and is reciprocal for any $\_k$ (see Ref.~\cite[\textsection~4]{suppl}). Although, the analogous finite crystal can break reciprocity, the effect, as a rule, is weak. Finally, a photonic crystal with three   layers per unit cell satisfies the necessary condition of spectral nonreciprocity (see Ref.~\cite[\textsection~4]{suppl}). It is possible to  mark out two possible scenarios. %The first scenario implies the use of birefringent materials~\cite{figotin_nonreciprocal_2001} and is not beneficial for us since natural birefringence is typically as low as $10^{-2} - 10^{-1}$~\cite[p.~176]{fowles_introduction_1989}, yielding optically large photonic crystal slab. 
%In the first scenario, the unit cell incorporates two different dielectric layers and single Weyl semimetal layer. This scenario   is examined in Ref.~\cite[\textsection~8]{suppl}. The second scenario, exhibiting the best characteristics in terms of the crystal size and implying two Weyl layers separated by a single dielectric spacing, is considered below.  

Figure~\ref{fig2}a depicts  the geometry of the proposed Voigt isolator. The unit cell comprises two semimetal layers with opposite in-plane Weyl nodes separations $ \_b$ and a dielectric layer made of fused silica with  permittivity  dispersion~$\E_2(\omega)$ measured in~\cite{malitson_interspecimen_1965}. %The Weyl semimetal properties are equivalent to those in Fig.~\ref{fig1}a.
The  Weyl layers have the same thickness of $x_1=x_3=480$~nm, while the thickness of dielectric layer is $x_2=1.44~\mu$m.  The total unit cell size is $\Lambda=x_1+x_2+x_3=2.4~\mu$m. We consider off-normal incidence since the crystal is reciprocal for normal incidence (see Supplemental Materials~\cite[\textsection~3]{suppl}).  
%As shown in Ref.~\cite[\textsection~4]{suppl}, the necessary condition of spectral nonreciprocity implies the incident wavevector to have $k_z=\beta$ component (the crystal is always reciprocal for normal incidence). 
Using the transfer matrix method~\cite{yeh_electromagnetic_1977}, we calculate the band structure of an \textit{infinite} photonic crystal  and the transmission spectrum of a \textit{finite} structure (see more details in Supplemental Materials~\cite[\textsection~4]{suppl}). Figure~\ref{fig2}b depicts the projected band structure  for an infinite crystal with respect to the $k_z=\beta$ component of incident wavevectors,  where the gray  and white  regions illustrate the allowed and forbidden bands, respectively. % As it is demonstrated in Ref.~\cite[\textsection~5]{suppl} based on the symmetry considerations, wave propagation with wavenumber $-\_k= -k_x \_x_0 - \beta\_z_0 $ (along Port~2 shown in Fig.~\ref{fig2}a) is equivalent to that with $k_x \_x_0 - \beta\_z_0 $ (along Port~4). Thus, the half of the plot in Fig.~\ref{fig2}b with negative values of $\beta$ can be attributed to the backward propagation with  $-\_k$ (see the red arrow  in Fig.~\ref{fig2}a).  
Such projected band structure is highly asymmetric with respect to $\beta$, which is expected to lead to nonreciprocal transmission for a finite structure.     Figure~\ref{fig2}c depicts the transmittance spectra of a finite structure with three unit cells for the two opposite illuminations with  wavevectors $\_k$ and $-\_k$  at an incident angle $\theta=45.6^\circ$.
%Next, we choose $\beta_0=0.33\cdot 2\pi/\Lambda$ providing the highest asymmetry contrast (see the blue lines in Fig.~\ref{fig2}b) and plot in Fig.~\ref{fig2}c the transmittances for the two opposite illuminations versus frequency. The data were calculated for the \textit{finite} photonic crystal including three unit cells (see derivations in Ref.~\cite[\textsection~7]{suppl}), i.e. in total nine material layers.  
The isolation ratio reaches 31.5~dB with an insertion loss of $1.2$~dB at $\omega=0.4619 \times 2\pi c/\Lambda$ ($\lambda=5.2~\mu$m). We note  that this frequency is below the bulk plasmon frequency, meaning that the high transmission in the forward direction is analogous to the effect of  the wave tunnelling in single-negative materials~\cite{jiang_properties_2004}.
Remarkably, the high isolation ratio is achieved with the total thickness $3\Lambda=7.2~\mu{\rm m}=1.38~\lambda$. Plots of reflectance spectra can be found in Supplemental Materials~\cite[\textsection~5]{suppl}. Interestingly, this device operates as an optical circulator with four ports shown in Fig.~\ref{fig2}a. The nonreciprocal wave propagation paths of this circulator are $1\rightarrow 2$, $2\rightarrow 3$, $3\rightarrow 4$, and $4\rightarrow 1$, where the numbers correspond to the ports. Similar functionality was previously achieved, e.g. in~\cite{kobayashi_microoptic_1980}, using conventional magneto-optical materials together with additional elements, such as quartz rotator, the Rochon prism, and mirrors. Our construction here is far simpler. 

Our designs above utilize the giant nonreciprocity associated with the bulk conductivity $\={{\sigma}}_{\rm B}$ of Weyl semimetals. Weyl semimetals also feature topological Fermi arc surface states when $\mathbf{b}$ has components parallel to the surface, such as in the Voigt geometry~\cite{belopolski2016}. Such surface states lead to a  surface conductivity $\={{\sigma}}_{\rm S}$, which  modifies the  boundary conditions for  electromagnetic waves:  $\hat{\_y} \cdot [\_H (x+0) - \_H(x-0)]= \hat{\_y} \cdot \={\sigma}_{\rm S}\cdot\_E (x) $. In our calculation, we neglect the surface term on the right side of this equation. This is because the relative importance of the bulk and surface terms is determined by the magnitudes of  $\={{\sigma}}_{\rm B}$ and $k\={{\sigma}}_{\rm S}$. %In the frequency and wavevector regime of isolation,   $k\={{\sigma}}^S \ll \={{\sigma}}^B$, thus the surface term is negligible.
We numerically verified that in the frequency and wavevector regime of isolation operation, for a typical surface conductivity ${{\sigma}}_{\rm S}\sim e^2/h$~\cite{chen_optical_2019},  $k\={{\sigma}}_{\rm S} \ll \={{\sigma}}_{\rm B}$ and the surface term $\hat{\_y} \cdot \={\sigma}_{\rm S}\cdot\_E (x) < 0.01\cdot H_y (x) $, thus is negligible.

%In the calculations above, we assumed the standard boundary conditions. However, at the Weyl semimetal interface in the configuration shown in~\ref{fig2}a, there excited surface states due to  Fermi arcs which modify the boundary conditions~\cite{chen_optical_2019}. In particular, the magnetic field at the interface becomes discontinuous $H_y (x+0) - H_y(x-0)= \_y_0 \cdot \={\sigma} \cdot \_E (x) $, where $\={\sigma}$ is the surface conductivity tensor. However, we have verified that for typical conductivity values of $q^2_{\rm e}/(2\pi \hbar)$~\cite{chen_optical_2019}, the contribution of the  
%surface states in the boundary conditions (calculated as the ratio of the right- and left-hand sides in the boundary condition above) does not exceed 1\% within the isolation region of the photonic crystal.  

Despite the explosive growth of research  on Weyl semimetals in condensed matter, its applications for photonics  are yet to be explored. The above two  examples of compact optical isolators clearly demonstrate that magnetic Weyl semimetals provide unprecedented material response that will open up new avenues for the design of nonreciprocal components. The absence of the external bias field and the subwavelength dimensions provide Weyl semimetal-based non-reciprocal optical components unique advantages as compared with  those based on  magneto-optical or time-modulation effects.  Similarly to the many  applications enabled by the discovery of graphene, its three-dimensional analogue, Weyl semimetal, may also lead to useful and novel photonic devices.

This work was supported in part by the Finnish Foundation for Technology Promotion and by the U.S. Air Force Office of Scientific Research Grant No. FA9550-18-1-0379.

\bibliographystyle{apsrev4-1}  % comment if want to have titles in the bibliography!
%\bibliography{Reference}% Produces the bibliography via BibTeX.


\end{document}